\documentclass{emulateapj}

\slugcomment{Accepted by AJ on 2011 July 21}

\shorttitle{Candidate Intermediate-Mass Black Hole in M54}
\shortauthors{Wrobel, Greene, \& Ho }

\begin{document}

\title{The Candidate Intermediate-Mass Black Hole in the Globular
  Cluster M54}

\author{J. M. Wrobel\altaffilmark{1},
        J. E. Greene\altaffilmark{2}, and
        L. C. Ho\altaffilmark{3}}

\altaffiltext{1}{National Radio Astronomy Observatory, P.O. Box O,
  Socorro, NM 87801, USA; jwrobel@nrao.edu}

\altaffiltext{2}{Department of Astronomy, University of Texas, Austin,
  TX 78712, USA; jgreene@astro.as.utexas.edu}

\altaffiltext{3}{The Observatories of the Carnegie Institution for
  Science, 813 Santa Barbara Street, Pasadena, CA 91101, USA;
  lho@obs.carnegiescience.edu}

\begin{abstract}
Ibata et al. reported evidence for density and kinematic cusps in the
Galactic globular cluster M54, possibly due to the presence of a
$9400~M_\odot$ black hole.  Radiative signatures of accretion onto
M54's candidate intermediate-mass black hole (IMBH) could bolster the
case for its existence.  Analysis of new {\em Chandra\/} and recent
{\em Hubble Space Telescope\/} astrometry rules out the X-ray
counterpart to the candidate IMBH suggested by Ibata et al.  If an
IMBH exists in M54, then it has an Eddington ratio of $L(0.3-8~{\rm
  keV}) / L({\rm Edd}) < 1.4 \times 10^{-10}$, more similar to that of
the candidate IMBH in M15 than that in G1.  From new imaging with the
NRAO Very Large Array, the luminosity of the candidate IMBH is
$L(8.5~{\rm GHz}) < 3.6 \times 10^{29}$~ergs~s$^{-1}$ (3 $\sigma$).
Two background active galaxies discovered toward M54 could serve as
probes of its intracluster medium.
\end{abstract}


\section{Motivation}\label{motivation}

Quantifying the space density of low-mass black holes (BHs) in the
local Universe has important implications for predicting gravity wave
signals \citep{hug09}, for understanding formation channels for seed
BHs \citep{vol08}, and for testing simulations of gravitational wave
recoil \citep{hol08}.  Moreover, establishing the existence or
break-down of scaling relations between central BHs and galaxies at
the low-mass end will help identify the physical processes driving the
relation between BH mass $M_{\rm BH}$ and stellar velocity dispersion
$\sigma_\ast$ \citep[e.g.,][]{gre10}.

Planned NIR telescopes in the 30 m class will eventually offer access
to BHs with $M_{\rm BH} \gtrsim 10^5~M_\odot$ in the local Universe
\citep{fer03,fer05}.  But lower-mass BHs will remain difficult to
find, as their observational signature on surrounding stars is limited
to a prohibitively small spatial scale beyond the Local Group.  We
refer to these systems as intermediate-mass black holes (IMBHs), since
they occupy the gap in mass between the well-studied stellar-mass BHs
with $M_{\rm BH} \sim 10~M_\odot$ and the well-established BHs with
$M_{\rm BH} \gtrsim 10^5~M_\odot$.

BH masses of order $10^4~M_\odot$ have been estimated for two globular
clusters, namely G1 in M31 \citep{geb02} and $\omega$~Cen in the
Galaxy \citep{noy10}.  Both estimates have been controversial.  For G1
the objections of \citet{bau03} were countered by \citet{geb05}.  For
$\omega$~Cen \citet{van10} provide a contradicting mass upper limit.

\citet{iba09} reported the detection of stellar density and kinematic
cusps in M54 (NGC\,6715), a globular cluster at a distance of 26.3 kpc
in the center of the Sagittarius dwarf galaxy
\citep{bel08}.\footnote{1\arcsec\, = 0.13 pc. The core, half-mass and
  limiting radii of M54 are 6.6\arcsec\, = 0.86 pc, 29.4\arcsec\, =
  3.8 pc and 6.3\arcmin\, = 49 pc, respectively
  \citep{har96,iba09,sol10}.}  Ibata et al. interpreted their findings
as possible evidence for a central BH with $M_{\rm BH} \sim
9400~M_\odot$, similar to G1 and possibly similar to $\omega$~Cen.
But Ibata et al. also noted that the orbits of the cusp stars in M54
could have moderate radial anisotropies.  They demonstrated that the
strong density gradient in a cusp could impede isotropic orbits from
being established, contrary to the common view that isotropic orbits
always arise when cluster relaxation times are relatively short. If
such anisotropic orbits are present, they could mimic the signature of
a central IMBH.

The interpretation of M54's cusp data is thus ambiguous, and other
lines of evidence for, or against, an IMBH should be sought.
Simulations suggest a globular cluster with primordial binary stars
can achieve a ratio of core to half-mass radius of only 0.08 or less,
whereas larger ratios can be realized by adding an IMBH
\citep{tre07,fre07}.  For M54 this ratio is about 0.2 \citep{har96}.
Also, simulations of massive disk galaxies suggest that several
tidally-stripped cores of dwarf satellites could retain their IMBHs
\citep{bel10}.  This fits in with M54's location in the center of the
disrupting Sagittarius dwarf galaxy \citep{bel08}.

Radiative signatures of accretion onto M54's candidate IMBH could
bolster the case for its existence.  \citet{ram06a} detected several
{\em Chandra\/} sources within M54's half-mass radius and suggested,
based on the sources' X-ray luminosities and colors, that they were
cataclysmic variables or low-mass X-ray binaries.  However, they noted
that one possibly-blended source, X-ray Id 2, was located within
1\arcsec\, (0.13 pc) of the center of the stellar density, taken a
proxy for the center of mass.  This led \citet{iba09} to suggest that
X-ray Id 2 could be the counterpart to the candidate IMBH in M54.

In this paper, we compare recent subarcsecond astrometry for the
stellar density center \citep{gol10} to improved astrometry for X-ray
Id 2 \citep{eva10}, ruling out an X-ray counterpart to the candidate
IMBH.  We also present new photometry of M54 with the NRAO Very Large
Array (VLA)\footnote{Operated by the National Radio Astronomy
  Observatory, which is a facility of the National Science Foundation,
  operated under cooperative agreement by Associated Universities,
  Inc.}  \citep{tho80}, detecting neither the candidate IMBH nor any
stellar emitters within the cluster's half-mass radius.  Two X-ray
sources with radio counterparts are detected within M54's limiting
radiius.  The new X-ray and radio data are presented in \S~\ref{data}
and their implications are explored in \S~\ref{implications}.  A
summary and conclusions appear in \S~\ref{sumcon}.

\section{Data}\label{data}

\subsection{Chandra Source Catalog}\label{Chandra}

A cone search toward the stellar density center \citep{iba09} was made
using version 1.1\footnote{http://cxc.harvard.edu/csc/} of the {\em
  Chandra\/} Source Catalog \citep[CSC;][]{eva10}.  Table 1 gives the
broadband (0.3 - 8 keV) ACIS positions returned for X-ray Ids 1-6 of
\citet{ram06a}, along with the position error radii at the 95\%
confidence level that includes an absolute astrometric error.  The
faintest source reported by \citet{ram06a}, X-ray Id 7, was not
recovered from the CSC.  If the position error radius for the
\citet{ram06a} astrometry is dominated by the absolute accuracy of
0.6\arcsec\, at the 90\% confidence limit\footnote{
  http://cxc.harvard.edu/cal/ASPECT/celmon}, the \citet{ram06a}
positions for X-ray Ids 1-6 agree with those reported in Table 1.
X-ray Id 2 was noted as a possibly blended source by \citet{ram06a}.
This trait is confirmed from the broadband CSC data: X-ray Id 2 is
spatially resolved, with a deconvolved 1 $\sigma$ radius of
0.67$\pm$0.18\arcsec\, that is indicated in Figure 1.

\begin{deluxetable}{llclc}
\tabletypesize{\scriptsize}
\tablecolumns{5}
\tablewidth{0pc}
\tablecaption{Subarcsecond Astrometry of Cluster Components}\label{t1}
\tablehead{
\colhead{R.A.}      & \colhead{Decl.}     & \colhead{Error}     &
\colhead{}          & \colhead{}          \\
\colhead{(J2000)}   & \colhead{(J2000)}   & \colhead{(\arcsec)} &
\colhead{Component} & \colhead{Ref.}      \\
\colhead{(1)}       & \colhead{(2)}       & \colhead{(3)}       &
\colhead{(4)}       & \colhead{(5)}       }
\startdata
18 55 02.45& -30 28 57.6& 0.56 & X-ray Id 4    & 1\\
18 55 02.77& -30 28 53.1& 0.59 & X-ray Id 6    & 1\\
18 55 02.95& -30 28 45.1& 0.43 & X-ray Id 1    & 1\\
18 55 03.46& -30 28 47.6& 0.46 & X-ray Id 2    & 1\\
18 55 03.345&-30 28 47.1& 0.72 & Stellar density center& 2\\
18 55 03.33& -30 28 47.5& 0.2  & Stellar density center& 3\\
18 55 03.65& -30 28 41.0& 0.62 & X-ray Id 5    & 1\\
18 55 03.89& -30 28 38.1& 0.53 & X-ray Id 3    & 1\\
\enddata 
\tablecomments{
Cols.~(1) and (2): Component position.  Units of right ascension are
hours, minutes, and seconds, and units of declination are degrees,
arcminutes, and arcseconds.
Col.~(3): Radius of error circle at 95\% confidence level.
Col.~(4): Component name.
Col.~(5): Reference.}
\tablerefs{(1) this work; (2) Ibata et al. 2009; (3) Goldsbury et al. 2010.}
\end{deluxetable}

\begin{figure}[t]
\plotone{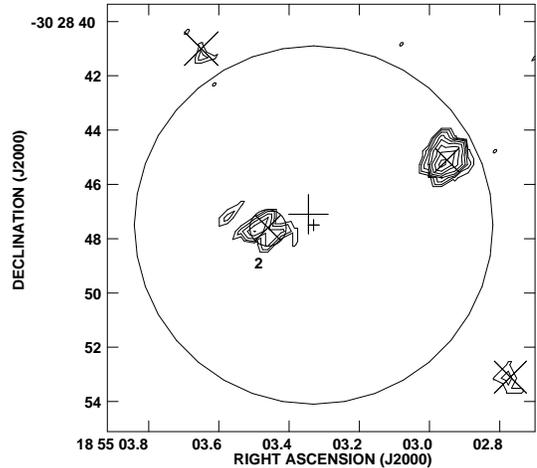}
\caption{{\em Chandra\/} broadband (0.3 - 8 keV) ACIS image of the
  central 16\arcsec\, (2.1 pc) of the globular cluster M54.  Contours
  are separated by the square root of 2 and start at
  $2.0\times10^{-6}$ photons~cm$^{-2}$~s$^{-1}$.  The large circle
  shows M54's core radius (6.6\arcsec\, = 0.86 pc) and its origin is
  at the stellar density center from \citet{gol10} marked with a small
  plus sign.  The stellar density center from \citet{iba09} is marked
  with a large plus sign.  The crosses mark the CSC version 1.1
  locations of the innermost X-ray Ids given in Table 1.  The sizes of
  the plus signs and the crosses convey position errors at the 95\%
  confidence level.  X-ray Id 2, labelled, is spatially resolved and
  its deconvolved 1 $\sigma$ radius of 0.67\arcsec\, is indicated by
  the small circle.}\label{f1}
\end{figure}

\subsection{VLA Imaging}\label{VLA}

The C configuration of the VLA was used, under proposal code AH1001,
to observe M54 near transit on UT 2009 July 12, 16, 23 and 2009 August
13.  The {\em a priori\/} pointing position for M54 was placed
10\arcsec\, North of the stellar density center \citep{iba09} to avoid
any phase-center artifacts.  Observations were made assuming a
coordinate equinox of 2000 and were phase referenced to the calibrator
J1845-2852 at an assumed position of $\alpha(J2000) = 18^{h} 45^{m}
51\fs3683$ and $\delta(J2000) = -28\arcdeg 52\arcmin 40\farcs276$ with
one-dimensional errors at 1 $\sigma$ better than 1
mas.\footnote{http://www.aoc.nrao.edu/software/sched/catalogs/sources.vlba}
The switching time between M54 and J1845-2852 was 240~s, while the
switching angle was 2.6\arcdeg.  A net exposure time of 11,500~s was
achieved for M54.  The center frequency was 8.4601~GHz, abbreviated as
8.5~GHz hereafter.  Data were acquired with a bandwidth of 100~MHz for
each circular polarization.  Observations of 3C\,48 were used to set
the amplitude scale to an accuracy of about 3\%.  All but two or three
of 27 antennas provided data of acceptable quality, with most data
loss attributable to EVLA retrofitting activities.

The data were calibrated using the 2009 December 31 release of the
NRAO AIPS software.  No polarization calibration or self-calibrations
were performed.  After calibration, each day's visibility data for M54
were concatenated.  The AIPS task {\tt imagr} was applied to the
concatenated data to form and deconvolve a naturally-weighted image of
the Stokes $I\/$ emission.  This image was corrected for primary-beam
attenuation with the AIPS task {\tt pbcor} and is shown in Figure 2.
From Figure 2, a 3 $\sigma$ upper limit of 51~$\mu$Jy~beam$^{-1}$ can
be placed on any 8.5-GHz emitters within the cluster's half-mass
radius.

\begin{figure}[t]
\plotone{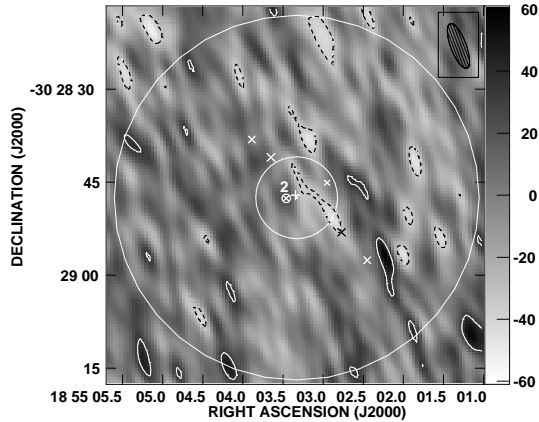}
\caption{VLA image of Stokes $I\/$ emission at 8.5~GHz of the central
  1\arcmin\, (7.8 pc) of the globular cluster M54.  Natural weighting
  was used, giving an rms noise of 17~$\mu$Jy~beam$^{-1}$ (1 $\sigma$)
  and beam dimensions at FWHM of 7.64\arcsec\, $\times$ 2.49\arcsec\,
  with elongation PA = 20.6\arcdeg\, (hatched ellipse).  Contours are
  at -6, -4, -2, 2, 4, and 6 times 1 $\sigma$.  Negative contours are
  dashed and positive ones are solid.  Linear grey scale spans
  -60~$\mu$Jy~beam$^{-1}$ to 60~$\mu$Jy~beam$^{-1}$.  Concentric
  circles show M54's core radius (6.6\arcsec\, = 0.86 pc) and
  half-mass radius (29.4\arcsec\, = 3.8 pc), and their origins
  coincide with the stellar density center from \citet{gol10}.  The
  symbol sizes and meanings are as for Fig.~1.}\label{f2}
\end{figure}

\subsection{Validating the Chandra Astromety}\label{validate}

Two {\em Chandra\/} sources \citep{ram06b} within M54's limiting
radius have counterparts at 8.5 GHz.  The VLA sources are
J185500.12-303049.7 and J185510.68-302650.9, with integrated flux
densities of 1.23$\pm$0.07 mJy and 3.78$\pm$0.13 mJy, respectively.
Each has a diameter less than 3\arcsec\, and a position error radius,
dominated by the phase-referencing strategies, of 0.2\arcsec.  Cone
searches toward these two using CSC version 1.1 return broadband ACIS
counterparts J185500.13-303049.4 with an error radius of 0.53\arcsec\,
and J185510.67-302651.1 with an error radius of 0.48\arcsec; each has
a diameter less than 1\arcsec.  The VLA and {\em Chandra\/} positions
agree within their combined errors, offering an independent validation
of the X-ray astrometry.

\section{Implications}\label{implications}

\subsection{Candidate IMBH}\label{imbh}

The energy equipartion argument of \citet{poo06} implies that the
candidate IMBH in M54, with mass $M_{\rm BH} \sim 9400~M_\odot$
\citep{iba09}, should coincide with the cluster's center of mass to
within 0.18\arcsec\, (0.02 pc).  Using the center of the stellar
density as a proxy for the center of mass, the position of the
candidate IMBH will be indistinguishable from those for the stellar
density center reported in Table 1.  Under these assumptions, the best
position for the candidate IMBH is that provided by \citet{gol10}.

Figure 1 shows the locations of the innermost cluster components in
Table 1.  The positions for the stellar density center \citep{gol10}
and X-ray Id 2 are offset by 1.7\arcsec\, (0.22 pc).  The quadratic
sum of the 1 $\sigma$ error in the stellar astrometry (0.1\arcsec) and
the X-ray astrometry (0.23\arcsec) is about 0.25\arcsec.  This leads
to a normalized offset of $1.7\arcsec / 0.25\arcsec \sim 6.8$, larger
than the upper limit of 3 signifying positional coincidence
\citep{con95}.  Thus X-ray Id 2 coincides neither with the stellar
density center nor, by the reasoning above, with the candidate IMBH.
Although X-ray Id 2 is spatially resolved, its radial extent does not
significantly alter this conclusion.  This astrometric result for
X-ray Id 2 is consistent with its luminosity and colors indicating a
possible blend of cataclysmic variables or low-mass X-ray binaries
\citep{ram06a}.

Adopting the unabsorbed flux limit and power-law index quoted in
\citet{ram06a}, the candidate IMBH in M54 has a luminosity of
$L(0.3-8~{\rm keV}) < 1.7 \times 10^{32}$~ergs~s$^{-1}$ and an
Eddington ratio of $L(0.3-8~{\rm keV}) / L({\rm Edd}) < 1.4 \times
10^{-10}$.  This latter limit is similar to the limit of
$L(0.2-10~{\rm keV}) / L({\rm Edd}) < 2.2 \times 10^{-9}$ reported by
\citet{ho03} for the candidate IMBH in M15.  In constrast, the
candidate IMBH in G1 has a considerably higher Eddington ratio of
$L(0.3-7~{\rm keV}) / L({\rm Edd}) \sim 10^{-6}$ \citep{kon10}.  If an
IMBH exists in M54, its accretion state resembles that of the
candidate IMBH in M15 rather than that of G1.  Interestingly, emerging
evidence suggests that accreting BHs in active galactic nuclei might
divide into ``quiescent'' and ``low'' states near an Eddington ratio
of about $10^{-6}$ \citep[e.g.,][and references therein]{ho09,yua09}.

For accreting BHs, the fraction of the accretion power that emerges in
the radio is a strongly increasing function of BH mass
\citep{mer03,fal04}.  This empirical relation among X-ray and radio
luminosity and BH mass was recast for the IMBH regime in equation (1)
of \citet{mac04}.  From that relation, for the same X-ray luminosity,
a $10^4~M_\odot$ BH will have $\sim 200$ times the radio flux density
of a $10~M_\odot$ BH.  Such a difference is well beyond the
factor-of-eight uncertainty in predicting the radio flux density from
the empirical relation.  In the case of G1, \citet{ulv07} successfully
detected the flux density predicted from the dynamically measured BH
mass and the X-ray luminosity.  That finding effectively ruled out the
possibility that the X-rays in G1 emanate from a stellar-mass BH.
However, improvements in the radio versus X-ray localizations of G1
could invalidate this inference \citep{kon10}.

Using the upper limit to the X-ray luminosity at the stellar density
center of M54 and applying equation (1) of \citet{mac04}, the
predicted flux density near 5 GHz is less than
20-1000~$\mu$Jy~beam$^{-1}$ for a $9400~M_\odot$ BH and
0.1-5~$\mu$Jy~beam$^{-1}$ for a $10~M_\odot$ BH.  From Figure 2, the
8.5-GHz flux density is less than 51~$\mu$Jy~beam$^{-1}$,
corresponding to a luminosity of $L(8.5~{\rm GHz}) < 3.6 \times
10^{29}$~ergs~s$^{-1}$.  Detection of a 10~$\mu$Jy source at the
stellar density center of M54 could help distinquish between these two
cases.  Detecting such a faint source is now feasible \citep{per11}
and would help validate the physically plausible radio predictions
advocated by \citet{mac08} and \citet{mac10}.  Also, M54 is more
distant than any other Galactic globular cluster surveyed deeply for
radio emission \citep{joh91,der06,bas08,cse10,lu11,mil11}, which
serves to reduce contamination from low-luminosity stellar emitters.

\subsection{Stellar Populations}\label{stellar}

The stellar population of M54 is old and metal poor
\citep[e.g.,][]{sie07}.  The {\em Chandra\/} sources reported by
\citet{ram06a} and confirmed in Table 1 are typical for Galactic
globular clusters \citep[e.g.,][]{hei11}.  From Figure 2, any stellar
emitters within the cluster's half-mass radius have a luminosity of
$L(8.5~{\rm GHz}) < 3.6 \times 10^{29}$~ergs~s$^{-1}$ (3 $\sigma$).
At this level, analogs of AC211 and M15 X-2 \citep[][and references
  therein]{mil11} would escape detection.  Planetary nebula are known
to be associated with the Sagittarius dwarf \citep{zil06} but none
have been reported for M54 itself.  An analog of K648 in M15
\citep{joh91} would be a mJy-level source at M54.  The mJy-level
sources within M54's limiting radius have X-ray counterparts,
inconsistent with being planetary nebulae.

\subsection{Intracluster Medium}\label{medium}

Because of M54's low Galactic latitude and distance
\citep{har96,bel08}, it is difficult to constrain its intracluster
medium and thus the fuel available to the candidate IMBH.
\citet{bur99} report an upper limit of $200~M_\odot$ (3 $\sigma$) for
the \ion{H}{1} averaged over 20 km~s$^{-1}$ toward the inner
21\arcmin\, = 160 pc of the Sagittarius dwarf.  This \ion{H}{1} mass
limit encompasses M54 in space, velocity and $\sigma_\ast$.  Two X-ray
sources \citep{ram06b} with 8.5-GHz counterparts each lie about
2\arcmin\,= 16 pc from the cluster center and within M54's limiting
radiius.  Their X-ray and radio properties strongly suggest that they
are active galaxies beyond M54.  As such, they could serve as probes
of M54's intracluster medium through studies of their Faraday rotation
measures.

\section{Summary and Conclusions}\label{sumcon}

Radiative signatures of accretion onto M54's candidate IMBH, with mass
$\sim 9400~M_\odot$ \citep{iba09}, could strengthen the case for its
existence.  Comparison of new X-ray and recent optical astrometry does
not support the suggestion by \citet{iba09} that X-ray Id 2 is the
counterpart to the candidate IMBH.  Rather, the Eddington ratio of the
candidate IMBH is constrained to be less than $1.4 \times 10^{-10}$,
implying a very different accretion state than that of G1's candidate
IMBH.  The 8.5-GHz luminosity of M54's candidate IMBH is less than
$3.6 \times 10^{29}$~ergs~s$^{-1}$.  Two background active galaxies
behind M54 could serve as probes of the fuel available to the
candidate IMBH.

\acknowledgments 
This research has made use of data obtained from the {\em Chandra\/}
Source Catalog, provided by the {\em Chandra\/} X-ray Center (CXC) as
part of the {\em Chandra\/} Data Archive.

{\it Facilities:} \facility{{\em Chandra}}, \facility{VLA}.

\end{document}